\documentclass[final]{svjour2} 
\usepackage{graphicx}
\usepackage{rotating}
\usepackage{amssymb}
\usepackage{mathptmx}
\usepackage[numbers]{natbib}
\makeatletter
\journalname{Journal of Low Temperature Physics}

\bibpunct{}{}{,}{s}{}{,}
\begin{document}

\newcommand{\hdblarrow}{H\makebox[0.9ex][l]{$\downdownarrows$}-}

\title{AC bias characterization of low noise bolometers for SAFARI using an
Open-Loop Frequency Domain SQUID-based multiplexer operating between 1 and 5 MHz.}

\author{L. Gottardi$^1$ \and M. Bruijn$^1$ \and J.-R. Gao$^1$,$^2$
  \and R. den Hartog$^1$ \and R. Hijmering$^1$ \and H. Hoevers$^1$ \and
P. Khosropanah$^1$ \and P. de Korte$^1$ \and J. van der Kuur$^1$
\and M. Lindeman$^1$ \and M. Ridder$^1$}
\institute{1:SRON National Institute for Space Research,\\ 
Sorbonnelaan 2, 3584 CA Utrecht, The Netherlands\\
\\2: Kavli Institute of NanoScience, Faculty of Applied Sciences,\\
 Delft University of Technology,\\
Lorentzweg 1, 2628 CJ Delft, The Netherlands}

\date{XX.XX.2007}

\maketitle


\begin{abstract}
SRON is developing the Frequency Domain Multiplexing (FDM) read-out and the
ultra low NEP TES bolometers array for the infra-red spectrometer SAFARI on board of the Japanese space mission SPICA.
The FDM prototype of the instrument requires critical and
complex optimizations. For single  pixel characterization under AC
bias we are developing a simple FDM system working in the frequency
range from 1 to 5 MHz,  based on the open loop read-out
of a linearized two-stage SQUID amplifier and  high Q lithographic LC resonators. 
We describe the details of the experimental set-up
required to achieve  low power loading ($<1\mathrm{fW}$) and low noise (NEP
$\sim 10^{-19} \mathrm{W/\sqrt{Hz}}$) in the TES bolometers.   
We conclude the paper by  comparing  the performance of a $4\cdot
10^{-19} \mathrm{W/\sqrt{Hz}}$ TES bolometer measured under DC and AC
bias.

PACS numbers: 
\end{abstract}

\section{Introduction}

In this paper we describe a Frequency Domain Multiplexer meant for
laboratory tests and  
designed to increase the experimental throughput in the
characterization of TES bolometer array under AC bias. To
simultaneously measure a
large number of pixels a baseband feedback scheme
\cite{RolandLTD} is required. However, to perform a single pixel
characterization, each AC channel can be read-out sequentially in
time. In this way the SQUID amplifier dynamic range is less critical and  the
instrument and its electronics can be greatly simplified. 

We took into account the following
requirements while designing the multiplexer. It should allow the read-out
out of tenths of pixels biased at AC voltage in the frequency range from 1
to 5 MHz. Several pixels on the array should be biased at DC voltage
to allow direct comparison with the AC biased ones under  identical
experimental conditions. It should have easy-to-use read-out
electronics. The multiplexers will be used both with
ultra-low noise equivalent power (NEP) TES bolometers  and with high energy
resolving power x-ray microcalorimeters. The former require very low 
background power levels, which is achieved by means of light blocking
filters in the signal loom feedthrought and a  light-tight
assembly. Special care has been taken to design the magnetic
shielding and to improve the uniformity of  the applied magnetic field
across the array.  
In the second part of the paper we compare  the performance of a $4\cdot
10^{-19} \mathrm{W/\sqrt{Hz}}$ TES bolometer measured under DC and AC
bias at a  frequency of 1.3 MHz. In the AC bias case the bolometer is
measured with a multiplexer based on discrete
LC resonators and other circuit components used in our standard DC bias set-up.

\section{Overview of the  Open-Loop Frequency Domain Multiplexer}

\begin{figure}[htbp]
\centering
\includegraphics[width=0.75\textwidth,keepaspectratio]{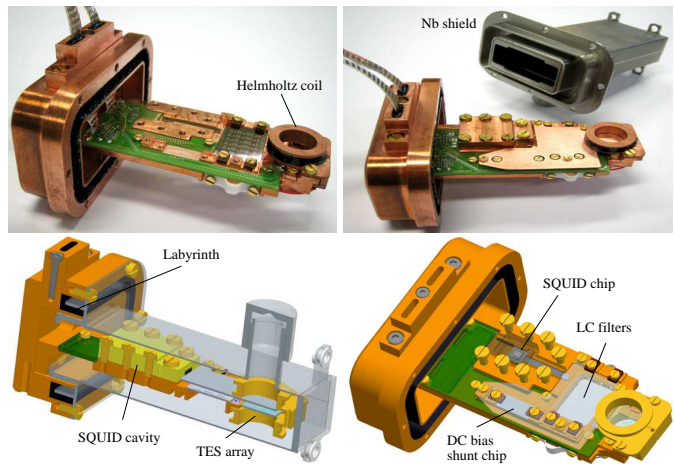}
\caption{ (Color online) Photograph and  CAD models of the
  Open Loop Frequency Multiplexer prototype. \label{OLFDMsetup}}
\end{figure}

The mechanical assembly of the Open Loop Frequency Domain Multiplexer (OL-FDM)
consists of a low magnetic impurity copper bracket fitted  
into a Nb can. 
The matching of the Nb can with the bracket lid is not vacuum-tight and  was  designed such
that it forms a labyrinth, which is filled with carbon loaded
epoxy on the copper lid side. In this configuration the Nb can provides both the
required magnetic and stray light shielding. A  photograph and a CAD
image of  the  OL-FDM set up is   shown  in Fig.~\ref{OLFDMsetup}. 
The electrical connections from the cold stage of the cooler to the
SQUID and TES bias circuit elements are achieved by means of
superconducting looms fed through a narrow 10 mm long channel filled with carbon loaded epoxy.
 The circuit PC board is designed to host 2 DC and 18 AC
 bias channels. The latter is achieved by using the lithografic high-Q
 LC resonators arrays developed at SRON \cite{MarcelLC}. The nominal inductance of
 the coil used in  each filter is $L=400 \mathrm{nH}$, while the
 capacitances C are designed such that the  frequencies
 $f_0=\frac{1}{2\pi\sqrt{LC}}$ are spread at a constant interval in the range from 1 to 5 MHz.

The SQUID amplifier chip  is placed in a
radiation shielding cavity whose inner side is coated with a 2 mm
thick radiation absorber made from carbon loaded epoxy with mixed  SiC grains
of size from $100 \mathrm{\mu m}$  to 1 mm \cite{epoxy1,epoxy2}. This
precaution was taken to minimize possible  loading of the bolometers
due to Josephson radiation, typically in the range of 4-8 GHz, emitted by the SQUID junctions. The
electrical connection from the SQUID chip to the LC filters is done by
means of Nb strip lines on a 20 mm long interconnection chip. These lines
act as  a low-pass filter with a calculated roll-off around 500 MHz. 
  
The perpendicular magnetic field of the TES array is controlled by
means of  a superconducting Helmholtz coil, which generates an
uniform field over the whole pixels array.

\section{Performance of the linearized SQUID amplifier}

A crucial component of the multiplexer described above is the SQUID
amplifier. We use a low noise two-stage  PTB SQUID current sensor with
on chip linearization, low input inductance ($\mathrm{L<3 nH}$) and low power dissipation ($\mathrm{P<20 nW}$)
\cite{OCFPTB}. The SQUID with on chip linearization \cite{MikkoOCF} has
  a larger dynamic  range with respect to the standard voltage sampled
  SQUID and guarantees a linear amplification of the AC biased
  bolometers signal.  We operate the SQUID in open loop. The output signal is
  amplified by a 20 MHz bandwidth, low input voltage noise,  commercial
  electronics (Magnicon B.V).
\begin{figure}[htbp]
\centering
\includegraphics[width=0.43\textwidth,keepaspectratio, angle=-90]{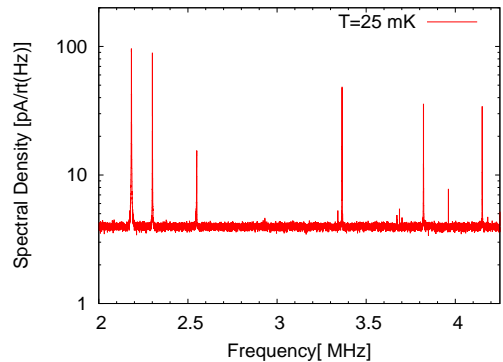}
\caption{Squid current noise with 6 coupled high-Q lithographic
  resonators. The Q-factor of the resonators ranges from 10000 to 20000. \label{SQUIDnoise}}
\end{figure}
To test the performance of the SQUID amplifer under loaded
  condition we coupled 6 high parallel LC lithographic
  resonators to its input coil. At temperature $T=25 \mathrm{mK}$ we
  measured a SQUID input current noise of $4
  \mathrm{pA/\sqrt{Hz}}$ (with $1/M_{in}=19.6\mu\mathrm{A/\Phi_0}$) over the whole interesting
  frequency range from 1 to 5 MHz (Fig.\ref{SQUIDnoise}). All the six
  resonators had Q factors larger than $10^4$.
 The SQUID operates in a linear regime for input current lower than
 $12 \mu \mathrm{A}$. 

\section{Characterization of a low noise bolometer at 1.3 MHz in a
  pilot set-up}
The experiment described in the following section was carried on in a
pilot set-up, described below, which is different from the one
previously presented.
The detector used in the experiment is  a low G TES
bolometers  with a transition temperature of  $T_C=78.5\,\mathrm{mK}$,
a  normal state  resistance  of 
$R_N=101\,\mathrm{m\Omega}$ and a calculated NEP of $2\cdot 10^{-19} \mathrm{W/\sqrt{Hz}}$. The sensor was previously characterized
under DC bias and showed a power plateau of $3.7\mathrm{fW}$ and a
dark NEP of $4.2\cdot 10^{-19} \mathrm{W/\sqrt{Hz}}$ \cite{RichardLTD}.    
We tested the bolometer under AC bias at a frequency of 1.3 MHz in a
three pixels FDM configuration, where two similar pixels were connected
to the other LC  resonators tuned  at 2.3 and 4.2 MHz respectively. The latter
pixels were not biased.
The TES array carrying the three pixels is mounted into the
light-tight box  used in the DC bias experiment \cite{RichardLTD} and the connections to the resonators were done by means of relatively long twisted pairs. 
  
The three LC  resonators consist  of  hybrid
filters made of  lithographic Nb-film  coils  and commercial
high-Q NP0  SMD capacitors with $C_1=22 \, \mathrm{nF}$, $C_2=4.4
\, \mathrm{nF}$, $C_1=2.2 \, \mathrm{nF}$ respectively.  
 For the read-out of the TES current we  use a NIST  SQUID arrays consisting of a  series of 100  dc-SQUID with
 input-feedback  coil  turns  ratio  of  3:1,  and  input  inductance
 $L_{in}=70 \, \mathrm{nH}$.   The   input   current   noise   is   $\sim
 3 \, \mathrm{pA/\sqrt{Hz}}$ at  $T<1\, \mathrm{K}$.  The SQUID  amplifier is
 operated  in open loop  mode using commercial Magnicon electronics.
The resonant  frequency of the circuit is defined by the  capacitors C
and  the  total  inductance $L_{tot}=L_{in}+L+L_{stray}$. From the measured resonant frequency and the filter capacitance value
reported above we get $L_{tot}=0.64\,\mu \mathrm{H}$.  

\begin{figure}[htbp]
\centering
\includegraphics[width=1\textwidth,keepaspectratio]{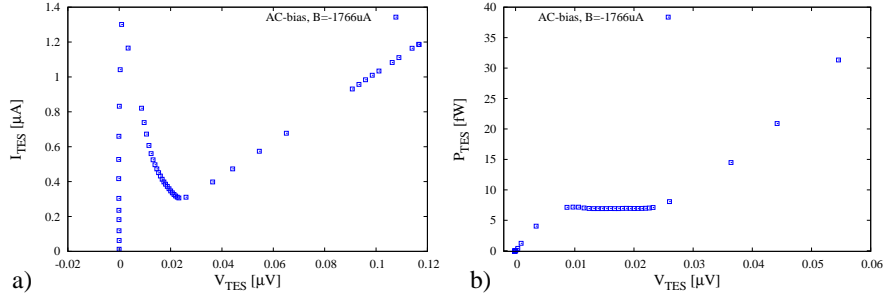}
\caption{ Current-to-voltage {\bf a.} and the power-to-voltage {\bf b.}
characteristics of the TES bolometer measured at an  AC bias frequency
of 1.3 MHz. \label{IVPV}}
\end{figure}

In Fig.\ref{IVPV} we show the current-to-voltage and the power-to-voltage
characteristics of the TES bolometer under test. We observe  a power
plateau at 7 fW at a bath temperature $T_{bath}=30 \mathrm{mK}$  and
when applying a magnetic field in order to cancel any residual fields
perpendicular to the TES. A lower power plateau of  3.7 fW was
observed with the same pixel in the
measurements performed at DC bias voltage under identical condition,
but in a different  cooler \cite{PouryaLTD}. 
The AC bias configuration performs even better than the DC bias case
in terms of power loading into the detector probably due  
to a better EMI filtering of the input bias lines.   

\begin{figure}[htbp]
\centering
\includegraphics[width=1.\textwidth,keepaspectratio]{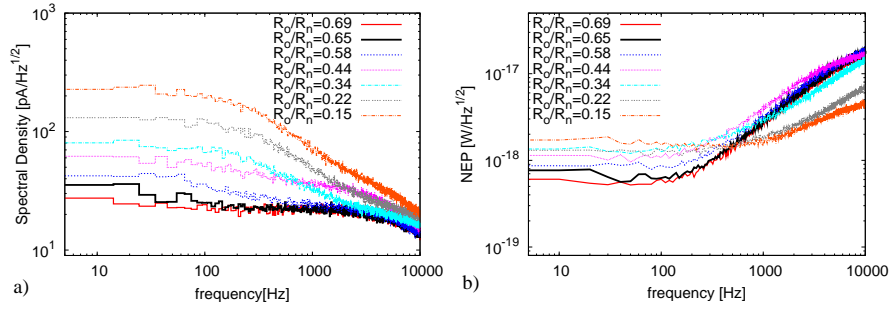}
\caption{ TES current noise {\bf a.} and dark NEP {\bf b.}
at different TES resistances  for the bolometer measured at an  AC
bias frequency of 1.3 MHz. \label{noisenep}}
\end{figure}

We further characterized the detector under AC bias by measuring 
 the noise and the complex impedance as a function of the TES resistance.  
A simple two body model \cite{YohLTD12}  is used to fit the impedance
data and to derive
the parameters, like $\beta$,  loop gain $L_o=P_o\alpha/GT$ and time constant $\tau_o=C/G$, needed to estimate the detector
responsivity. A distributed model would provide a more accurate
estimation of those parameters \cite{PouryaLTD}. In Fig.\ref{noisenep} we plot the TES current noise and
the dark NEP  for several TES resistances. 

\begin{figure}[htbp]
\centering
\includegraphics[width=0.6\textwidth,keepaspectratio]{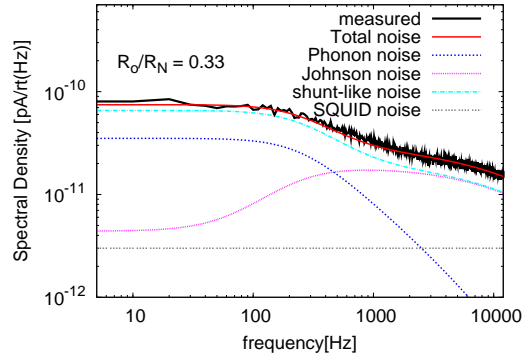}
\caption{ (Color on line) Measured and modelled current noise for the
  TES bolometer at a resistance $R/R_N=0.33$. \label{noisemodel}}
\end{figure}

At bias points high in the transition ($R/R_N>60\%$) an $\mathrm{NEP}=(5.2\pm
0.6)\cdot 10^{-19} \mathrm{W/\sqrt{Hz}}$ is observed. The uncertainty
in the NEP estimation is due to calibration errors. 
We observed a deterioration of
the dark NEP at low TES resistances due to excess noise at low
frequency. The detector noise can be modeled  using the parameter
obtained from the impedance data. 
We can fully explain the measured noise at any bias point in the transition assuming a voltage noise $\sqrt{S_V}\sim 2
\mathrm{pV/\sqrt{Hz}}$ in series with the TES. 
The current noise generated by the
voltage source  has the same signature of a
thermal
noise from the shunt resistance used in the DC bias case. This is
clearly seen in Fig.\ref{noisemodel}, where the different noise
contributions, including the shunt-like noise, are overplotted to the
measured noise spectrum. 

From the quality factor Q of the
resonance measured with the TES superconducting we infer a total
resistance of the read-out circuit of $r=\frac{1}{\omega_O C Q}=14.9
\pm 0.3 \, \mathrm{m\Omega}$, where $\omega_o=2\pi \cdot 1.33 \mathrm{MHz}$, $C=22
\mathrm{nF}$ and $Q=346 \pm 6$. Such a small resistance should be at a
thermodynamic temperature of about 4 K to generate the voltage noise level
needed to explain the measured TES current noise. The excess
noise observed  is  likely to be due to thermal noise leaking from the  channel 2 of the FDM
system, whose resonator was not functioning properly and had a poor quality factor.

\section{Conclusion}
We are developing a  Frequency Domain Multiplexer 
to increase the experimental throughput in the characterization of TES bolometer array under AC bias. 

The multiplexer is based on the open loop read-out
of a linearized two-stage SQUID amplifier and  high Q lithographic LC
resonators and is designed to work   in the frequency
range from 1 to 5 MHz,  
We describe the details of the experimental set-up
required to achieve  low power loading ($<1\mathrm{fW}$) and low noise (NEP
$\sim 10^{-19} \mathrm{W/\sqrt{Hz}}$) in the TES bolometers. The first
results are expected soon.  
In a pilot experiment performed with a multiplexer obtaind by adapting
our standard DC bias set-up, we measured a dark  $\mathrm{NEP}=(5.2\pm
0.6)\cdot 10^{-19} \mathrm{W/\sqrt{Hz}}$ using a low noise TES bolometer
previously characterized under DC bias. We observed a deterioration of
the dark NEP for low TES resistances due to excess noise at low
frequency. The excess noise is consistent with a voltage noise source
in series with the TES as large as $\sqrt{S_V}\sim 2
\mathrm{pV/\sqrt{Hz}}$ likely to be due to thermal noise leaking from
another  channel of the FDM system.           

\begin{acknowledgements}

We thank Manuela Popescu and Martijn Schoemans for their precious
technical help. 

\end{acknowledgements}


\end{document}